\documentclass[aps,preprint,showpacs]{revtex4}
\usepackage{graphicx}
\usepackage{dcolumn}
\usepackage{bm}

\newcommand{\tauiso}{{\mbox{\boldmath $\tau$}}}

\begin{document}
\title{Exclusive $K^+$ production in proton-nucleus collisions}
\author{R. Shyam\footnote{On leave from Saha Institute of Nuclear Physics,
Calcutta, India}, H. Lenske, and U. Mosel}
\affiliation{Institute f\"ur Theoretische Physik, Universit\"at Giessen,
D-35292 Giessen, Germany}

\date{\today}

\begin{abstract}
The exclusive $K^+$ meson production in a proton-nucleus collision, 
leading to two body final states, is investigated in a fully covariant
two-nucleon model based on the effective Lagrangian picture.  
The explicit kaon production vertex is described via creation,
propagation and decay into relevant channel of $N^*$(1650), $N^*$(1710)
and $N^*$(1720) intermediate baryonic states in the initial collision
of the projectile nucleon with one of its target counterparts which
is modeled by the one-pion exchange process. The calculated cross
sections show strong sensitivity to the medium effects on pion 
propagator and to the final hypernuclear state excited in the reaction.  
\end{abstract}
\maketitle
\section{Introduction}
$\Lambda$ hypernuclei have been studied extensively by stopped
as well as in-flight $(K^-,\pi^-)$
reaction~\cite{chr89} and also by $(\pi^+,K^+)$ reaction
\cite{pil91,hot01}. The kinematical properties
of the $(K^-,\pi^-)$ reaction allow only a small momentum 
transfer to the nucleus, thus there is a large probability of 
populating $\Lambda$-substitutional states 
in the residual hypernucleus. On the other hand, in the $(\pi^+,K^+)$
reaction the momentum transfer is larger than the nuclear Fermi momentum.
Therefore, this reaction can populate configurations with an outer 
neutron hole and a $\Lambda$ hyperon in a series of orbits covering all
bound states. During past years data on the hypernuclear
spectroscopy have been used extensively to extract information about
the hyperon-nucleon interaction (which could be quite different from
the nucleon-nucleon interaction) within a variety of theoretical
approaches (see, e.g.,~\cite{hiy00,kei00}).

Alternatively, $\Lambda$-hypernuclei can also be produced with high intensity 
proton beams via the $(p,K^+){_{\Lambda}}B$ reaction where the hypernucleus
${_{\Lambda}}B$ has the same neutrons and proton numbers as the target 
nucleus $A$.  First set of data for this reaction on deuterium
and helium targets, have already been reported in Ref.~\cite{kin98}. More
such measurements involving heavier targets are expected to be performed at
the COSY facility of the Forschungszentrum, J\"ulich~\cite{sch95}. 
The states of the hypernucleus
excited in the $(p,K^+)$ reaction may have a different
type of configuration as compared to those excited in the 
$(\pi^+,K^+)$ reaction. Thus a comparison of informations extracted from 
the study of two reactions is likely to lead to a better
understanding of the hypernuclear structure.

Theoretical studies of the $A(p,K^+){_{\Lambda}}B$ reaction 
are rather sparse and 
preliminary in nature~\cite{shi86,kom95,kri95}. They are 
based on two main approaches; the one-nucleon model (ONM) [Fig.~1(a)]
and the two-nucleon model (TNM) [Fig.~1(b) and 1(c)]. In the ONM 
the incident proton first scatters from the target nucleus and emits
a (off-shell) kaon and a $\Lambda$ hyperon. Subsequently, the kaon 
rescatters into its mass shell while the hyperon  gets 
captured into one of the (target) nuclear orbits. Thus 
there is only a single active nucleon (impulse approximation)
which carries the entire momentum transfer to the target nucleus
(about 1.0 GeV/c at the outgoing $K^+$ angle of 0$^\circ$ ).
This makes this model  extremely sensitive to details of the bound state
wave function at very large momenta where its magnitude is very small
leading to quite low cross sections. In the ONM
calculations of $(p,K^+)$ and also of $(p,\pi)$ reactions the distortion
effects in the incident and the outgoing channels have been found to be
quite important~\cite{kri95,mea79,coo82}). 

In the two-nucleon mechanism (TNM), on the other hand, the
kaon production proceeds via a collision of the projectile nucleon
with one of its target counterparts. This excites intermediate baryonic 
resonant states which decay into a kaon and a $\Lambda$ hyperon. 
The nucleon and the hyperon are captured into the respective nuclear 
orbits while the kaon rescatters into its mass shell. In this 
picture there are altogether three active bound state baryon wave functions
taking part in the reaction process allowing the large momentum
transfer to be shared among three baryons. Consequently, the 
sensitivity of the model is shifted from high momentum
parts of the bound state wave functions (not very well known)
to those at relatively lower momenta where they are rather
well known from $(e, e^\prime p)$ and $(\gamma, p)$ experiments and
are relatively larger.  This could lead to larger cross sections.
Moreover, in the TNM studies of other large
momentum transfer reactions like $(p,\pi)$, $(\gamma,\pi)$, and
$(\gamma,\eta)$ the distortion effects have been found to be less
pronounced~\cite{alo88,pet98}. This type of two-nucleon model has not
yet been applied to the study the $A(p,K^+){_{\Lambda}}B$ reaction.

All previous calculations~\cite{shi86,kom95,kri95} of the
$A(p,K^+){_{\Lambda}}B$ reaction, have been done within the non-relativistic
framework.  However, for processes involving momentum transfers of
typically 400 MeV/c or more, a non-relativistic treatment of corresponding
wave functions is questionable as the lower component of the
Dirac spinor is no longer suppressed in this region
(see, e.g, Ref.~\cite{shy95}).  
Furthermore, in the non-relativistic description one must
resolve the ambiguities of the pion-nucleon-nucleon vertex that result
from the non-relativistic reduction of the full covariant
pion-nucleon-nucleon vertex (see, e.g., \cite{alo88,shy95,bar69} for more
details). 

In this paper, we study the $A(p,K^+){_{\Lambda}}B$ reaction within a
fully covariant TNM by retaining the field theoretical structure of the 
interaction vertices and by treating the baryons as Dirac particles.
Similar to the model used in Ref.~\cite{shy99} to 
describe the $K^+$ production in elementary proton-proton ($pp$) collisions,
the initial interaction between the incoming proton and a bound nucleon of 
the target is described by the one pion exchange mechanism which makes
the dominant contribution to the elementary $pp K^+\Lambda$ process. It
should, however, be mentioned that some authors have used an alternative
approach for the elementary $pp K^+\Lambda$ reaction~\cite{kai99,gas00,lag91}
in which kaon exchange mechanism between the two initial state nucleons leads
to the $K^+$ production. We have not considered this mechanism here.

In our model, the initial state interaction of the incoming proton 
with a bound nucleon leads to 
$N^*$(1650)[$\frac{1}{2}^-$], $N^*$(1710)[$\frac{1}{2}^+$], and
$N^*$(1720)[$\frac{3}{2}^+$] baryonic resonance intermediate states  
which make predominant contributions to elementary $pp \to p K^+ \Lambda$
cross sections in the beam energy range from near threshold to 10 GeV
\cite{shy99}. Terms corresponding to interference among various resonance
excitations are included in the total reaction amplitude. We have ignored
the diagrams of type 1(a) since contributions of such processes are expected 
to be very small in comparison to those of the TNM. In this first
exploratory study our aim is primarily to establish the relativistic
TNM mechanism for the $(p,K^+)$ reaction. Therefore, to reduce the 
computational complications we have used plane waves to describe the
relative motions of incoming proton and outgoing kaon. 

In section II, we present the details of our formalism for 
calculating the amplitudes corresponding to the diagrams 
shown in Fig. 1b and 1c. The results of our calculations
performed for the $^{40}$Ca$(p,K^+)$$^{41}\!\!\!_\Lambda$Ca
reaction are presented and discussed in section III. Summary 
and conclusions of our work are given in section IV.

\section{Formalism}

The effective Lagrangian for the nucleon-nucleon-pion ($NN\pi$)
and $N^*$-nucleon-pion ($N^*N\pi$) vertices are given by
\begin{eqnarray}
{\cal L}_{NN\pi} & = & -\frac{g_{NN\pi}}{2m_N} {\bar{\Psi}} \gamma _5
                             {\gamma}_{\mu} \tauiso
                            \cdot (\partial ^\mu {\bf \Phi}_\pi) \Psi. \\
{\cal L}_{N_{1/2}^*N\pi} & = & -g_{N_{1/2}^*N\pi}
                          {\bar{\Psi}}_{N_{1/2}^*} i{\Gamma} \tauiso
                        {\bf \Phi}_\pi \Psi 
                          + {\rm h.c.},\\
{\cal L}_{N_{3/2}^*N\pi} & = & \frac{g_{N_{3/2}^*N\pi}}{m_\pi}
                         {\bar{\Psi}}_{\mu}^{N^*} \Gamma_\pi
                         {\tauiso} \cdot \partial ^{\mu}
                         {\bf \Phi}_\pi \Psi + {\rm h.c.}.
\end{eqnarray} 
We have used notations and conventions of Bjorken and Drell~\cite{bjo64}.
In Eq.~(1) $m_N$ denotes the nucleon mass. The operator $\Gamma$ ($\Gamma_\pi$)
is either $\gamma_5$ (unity) or unity ($\gamma_5$) depending upon the parity of
the resonance being even or odd, respectively. 
We have used a pseudovector (PV) coupling for the $NN\pi$ vertex and a
pseudoscalar (PS) one for the $N_{1/2}^*N\pi$ vertex which provides the
best description of the elementary $pp \to p\Lambda K^+$ data~\cite{shy99}.
The effective Lagrangians for the
resonance-hyperon-kaon vertices are written as 
\begin{eqnarray}
{\cal L}_{N_{1/2}^*\Lambda K^+} & = & -g_{N^*_{1/2}\Lambda K^+}
                          {\bar{\Psi}}_{N^*} {i\Gamma} \tauiso
                        {\bf \Phi}_{K^{+}} \Psi
                        + {\rm h.c.}.\\
{\cal L}_{N_{3/2}^*\Lambda K^+} & = & \frac{g_{N^*_{3/2}\Lambda K^+}}{m_{K^+}}
                         {\bar{\Psi}}_{\mu}^{N^*} \Gamma_\pi
                         {\tauiso} \cdot \partial ^{\mu}
                         {\bf \Phi}_{K^{+}} \Psi + {\rm h.c.}.
\end{eqnarray}
The operator $\Gamma$, in Eq.~(4), is defined in the same way as in
Eq.~(2). In Eqs.~(3) and (5) ${\Psi}_{\mu}^{N^*}$ is the vector spinor
for the spin-${3 \over 2}$ particle. Further discussions about the 
vertices involving such particles are given in Refs.~\cite{shy99,pen02}.
Signs and values of various coupling constants have been taken from
\cite{shy99} which are shown in Table I.
In Eqs.~(1)-(3), the $NN\pi$, $N_{1/2}^*N\pi$ and $N_{3/2}^*N\pi$
vertices are corrected for the off-shell 
effects by multiplying the corresponding coupling constants by
form factors of the same forms as used in Ref.~\cite{shy99}. The values
of the cut-off parameters appearing therein are taken to be 1.2 GeV
in all the cases. 

The propagators for pion and spin-${1 \over 2}$ and spin-${3 \over 2}$ 
intermediate resonances are given by  
\begin{eqnarray}
D_\pi(q) & = & {i \over {q^2 - m_\pi^2 -\Pi(q)}},\\
D_{N^*_{1/2}} (p) & = & i\left[ {p_\eta\gamma^\eta + m_{N^*_{1/2}}} \over
                       {p^2 - (m_{N^*_{1/2}}-i\Gamma_{N^*_{1/2}}/2)^2}
                         \right ],\\
D_{N^*_{3/2}}^{\mu \nu} (p) & = & -\frac{i(p\!\!\!/ + m_{N^*_{3/2}})}
                    {p^2 - (m_{N^*_{3/2}}-i\Gamma_{N^*_{3/2}}/2)^2}
                               \nonumber \\
                 &   &\times [g^{\mu \nu} - \frac{1}{3}\gamma^\mu \gamma^\nu
                              - \frac{2}{3m_{N^*_{3/2}}^2} p^\mu p^\nu
                              + \frac{1}{3m_{N^*_{3/2}}^2}
                                ( p^\mu \gamma^\nu - p^\nu \gamma^\mu )].
\end{eqnarray}
In Eq.~(6), $\Pi(q)$ is the (complex) pion self energy which accounts
for the medium effects on the propagation of the pion in the nucleus.
In Eqs.~(7) and (8), $\Gamma_{N^*}$ is the total width of the resonance
which is introduced in the denominator term to account for the
finite life time of the resonances for decays into various
channels. $\Gamma_{N^*}$ is a function of the center of mass momentum
of the decay channel, and it is taken to be the sum of the widths for pion
and rho decay (the other decay channels are considered only implicitly by
adding their branching ratios to that of the pion channel)~\cite{shy99}.
The medium corrections on the intermediate resonance widths has not been
included as we do not expect any major change in our results due to
these effects. As is pointed out in~\cite{lut03}, the medium correction
effects on  widths of the $s$- and $p$-wave resonances, which make the
dominant contribution to the cross sections being investigated here,
are quite small which can also be proved by following the method presented
in \cite{mal02}). On the other hand, any medium modification in the width of
the $d$-wave resonance is unlikely to alter our results significantly as
their contributions to the cross sections are negligibly small (see the
subsequent discussions). 

After having established the effective Lagrangians, coupling constants and
forms of the propagators, one can  write down, by following the well known
Feynman rules, the amplitudes for various graphs [Figs.~1b and 1c]
which can be evaluated numerically by using the techniques described
in Refs.~\cite{shy95,shy91}.  While writing the amplitudes 
the isospin part is treated separately which gives rise to a constant
factor for each graph. For example, the amplitude for graph 1b with
spin-${1 \over 2}$ baryonic resonance is given by,
\begin{eqnarray}
M_{1b}(N^*_{1/2}) & =& C_{iso}^{1b}\biggl(\frac{g_{NN\pi}}{2m_N}\biggr)
(g_{N_{1/2}^*N\pi}) (g_{N^*_{1/2}\Lambda K^+})
{\bar{\psi}}(p_2) \gamma _5 \gamma_\mu q^\mu \nonumber \\
& \times & \psi(p_1) D_{\pi}(q)
{\bar{\psi}}(p_\Lambda)\gamma_5 D_{N^*_{1/2}}(p_{N^*}) \gamma_5 \nonumber \\
& \times & \phi^{(-)*}_{K}(p_K^\prime, p_K)\psi^{(+)}_i(p_i^\prime, p_i),
\end{eqnarray}
where various momenta are as defined in Fig.~1b. In addition, $p_{N^*}$
is the momentum associated with the intermediate resonance and $p_\Lambda$
is that associated with the $\Lambda$ hyperon. The isospin factor 
$C_{iso}^{1b}$ is unity for both the graphs 1(b) and 1(c). The
functions $\psi$ are the four component (spin space) Dirac spinor in
momentum space~\cite{shy95,shy91}.
$\phi_{K}^{(-)*}(p_K^\prime, p_K)$ [$\psi_i^{(+)}(p_i^\prime, p_i)$]
is the wave function for the outgoing kaon [incoming proton] with
appropriate boundary conditions.

To get the $T$ matrix of the $(p,K^+)$ reaction, one has to integrate
the amplitude corresponding to each graph over all the independent
intermediate momenta subject to constraints imposed by the
momentum conservation at each vertex. For instance, for the amplitude
corresponding to Eq.~(6) the respective $T$ matrix is given by
\begin{eqnarray}
T_{1b}(N^*_{1/2}) & = & \int \frac{d^4p^\prime_i}{(2\pi)^4}
\int \frac{d^4p^\prime_K}{(2\pi)^4}
\int \frac{d^4p_\Lambda}{(2\pi)^4} \int
\frac{d^4p_2}{(2\pi)^4} \nonumber \\
& \times &
\delta(q - p_1 + p_2)\delta(p_{N^*} - p_{K} - p_\Lambda)\nonumber \\
& \times &
\delta(p_\Lambda - p_i^\prime - q + p_K^\prime) M_{1b}(N^*_{1/2}).
\end{eqnarray}

For the computation of the amplitudes we make use of the momentum space
Dirac equation in an external potential field~\cite{shy95,shy91}
\begin{eqnarray}
p\!\!\!/\psi(p) & = & m_N\psi(p) + F(p),
\end{eqnarray}
where
\begin{eqnarray}
F(p) & = & \delta(p_0 - E) \Biggl[\int d^3p^\prime V_s(-p^\prime)
\psi(p+p^\prime) \nonumber \\ & - &
 \gamma_0 \int d^3p^\prime V_v^0(-p^\prime) \psi(p+p^\prime) \Biggr] .
\end{eqnarray}
In Eq.~(12), $V_s$ and $V_v^0$ represent a scalar potential and
time-like component of a vector potential in the momentum space. Using
Eqs.~(11)-(12) the amplitudes can be reduced to a form which is suitable for
numerical computations.  In this study,
the incoming proton spinor $\psi^{(+)}_i(p_i^\prime, p_i)$ and the
outgoing kaon wave function $\phi^{(-)*}_K(p_K^\prime, p_K)$ are replaced
by their plane wave counterparts. This makes redundant the integrations over
variables $p_i^\prime$ and $p_K^\prime$ in Eq.~(10).
The final $T$ matrix ($T$) is obtained by summing the transition
matrices corresponding to all the graphs.

The differential cross section for the $(p,K^+)$ reaction is given by
\begin{eqnarray}
\frac{d\sigma}{d\Omega} & = & \frac{1}{(4\pi)^2}
\frac{m_pm_Am_B}{(E_{p_i} + E_A)^2} \frac{p_K}{p_i}
 \sum_{m_im_f} |T_{m_im_f}|^2,
\end{eqnarray}
where $E_{p_i}$ and $E_A$ are the total energies of incident
proton and the target nucleus, respectively while $m_p$, $m_A$ and $m_B$ are
the masses of the proton, and the target and residual nuclei, respectively.
The summation is carried out over initial ($m_i$) and final ($m_f$)
spin states. 

\section{Results and Discussions}

\subsection{The bound state spinors}

We have chosen the reaction $^{40}$Ca$(p,K^+)$$^{41}\!\!\!_\Lambda$Ca for
the first numerical application of our method.
The  spinors for the final bound hypernuclear state (corresponding to
momentum $p_\Lambda$) and for two intermediate nucleonic states
(corresponding to momenta $p_1$ and $p_2$) have been determined
by assuming them to be pure-single particle or single-hole states
with the core remaining inert. The quantum numbers of the two
intermediate states are taken to be the same. The spinors in the
momentum space are obtained by Fourier transformation of the
corresponding coordinate space spinors which are obtained by solving
the Dirac equation with potential fields consisting of an attractive
scalar part ($V_s$) and a vector part ($V_v$) with the Woods-Saxon
geometry. With a fixed set of the geometry parameters (reduced radii
$r_s$ and $r_v$ and diffusenesses $a_s$ and $a_v$), the depths of the
potentials were searched in order to reproduce the binding energies of
the particular state (the corresponding values are given in Table II).
The quantum numbers as well as the single baryon binding energies for final
hypernuclear states have been taken from the density dependent
relativistic hadron field (DDRH) theory predictions of Ref.~\cite{kei00},
which reproduce the corresponding experimental binding energies quite
well.
 
To have an idea of the relative strengths of the upper and lower
components of the Dirac spinors as a function of the transferred momentum,
we show, e.g., in Fig.~2(a) the $0p_{3/2}$ $\Lambda$ hyperon spinors in  
momentum space for the $^{41}\!\!\!_\Lambda$Ca hypernucleus. Similar
to the observation made in Ref.~\cite{shy95} for nucleon spinors, we
note that only for momenta $<$ 1.5 fm$^{-1}$, is the lower component of
the spinor substantially smaller than the upper component. In the region
of momentum transfer pertinent to to exclusive kaon production in 
proton-nucleus collisions, the magnitudes of the upper and lower components
are of the same order of magnitude. This clearly demonstrates that a 
fully relativistic approach is essential for an accurate description of 
processes.
 
The spinors calculated in this way provide a good description of the
experimental nucleon momentum distributions for various nucleon orbits
as is shown in Ref.~\cite{shy95}. In Fig. 2(b) we show the momentum
distribution of the $\Lambda$ hyperon in the $0p_{3/2}$ state in 
$^{41}\!\!\!_\Lambda$Ca. The momentum distribution is defined as 
$q^2[|F(q)|^2 + |G(q)|^2]$~\cite{fru84}. It is clear that the momentum
density of this hyperon shell is of the order of 10$^0$ at the
momentum of 0.35 GeV/c as compared to 10$^{-8}$ - 10$^{-9}$ at 
that of 1.0 GeV/c. Thus in the TNM where the sensitivity of the
process is shifted to lower momentum transfers, one expects to get a larger
cross section for the exclusive meson production reactions. 

\subsection{Strangeness production }

A crucial quantity needed in the calculations of the Kaon 
production amplitude [Eq.~(9)], is the pion self-energy, $\Pi(q)$,
which takes into account the medium effects on the intermediate pion
propagation. Since the energy and momentum carried by such a pion 
can be quite large (particularly at higher proton incident energies),
a calculation of $\Pi(q)$ within a relativistic approach is mandatory.
In our study we have adopted the approach of Ref.~\cite{dmi85} for 
this purpose where contributions from the particle-hole ($ph$) and
delta-hole ($\Delta h$) excitations are taken into account. The
self-energy has been renormalized by including the short-range repulsion
effects by introducing the constant Landau-Migdal parameter
$g^\prime$ which is taken to be the same for $ph-ph$ and $\Delta h-ph$
and $\Delta h-\Delta h$ correlations which is a common choice.
All the relevant formulas for this calculation are given in Ref.~\cite{shy95}.
The parameter $g^\prime$ is supposed to mock up the complicated
density dependent effective interaction between particles and holes in
the nuclear medium.  Most estimates give a value of $g^\prime$ between
0.5 - 0.7. Like previous studies of e.g., $(p,\pi)$~\cite{shy95}
and $(p,p^\prime \pi)$ \cite{jai88} reactions, $(p,K^+)$ cross sections too
are expected to show sensitivity to the parameter $g^\prime$.

In Fig.~3 we show, e.g., the pion self-energy calculated  
at zero energy ($q_0$ = 0) for several values of $g^\prime$
for a Fermi momentum of 230 MeV/c. $\Pi(q)$ is real and attractive
for this case.  We note that for $g^\prime$ = 0 the 
self energies are large; there magnitude decreases with increasing
value of the parameter $g^\prime$. It may be remarked here that 
the magnitudes of self-energies as shown in Fig.~3 (with $g^\prime$
=0.5 - 0.7) are in good agreement with those reported in Ref.~\cite{her92}
where calculations have been done within a relativistic mean field model.

Some remarks can already be made about the effect of the self-energy
correction (to the pion propagator) on various amplitudes. For the
diagramme 1(c), the energy ($q_0$) associated with the intermediate
pion is almost equal to that of the incident proton. Therefore, in the
absence of the self energy term, poles may appear in the momentum
integration for this diagramme. With the inclusion of self-energy
(which, in general, is complex at non-zero energies), the pole is
automatically removed from the real axis and the pole integration problem
disappears and the graph 1(c) becomes well behaved.

On the other hand, for the diagramme 1(b) where $q_0 \approx 0$,
the self-energy is real and attractive (as can be seen from Fig.~3). This
leads to a small denominator of the pion propagator leading to a
consequent increase in the contribution of this graph. Indeed, for 
magnitudes of the self-energies $\approx ({\bf q}^2 + m_\pi^2)$,
the integration in the $T$ matrix of graph 1(b) may even involve poles.
While the self-energy effects will give a considerable increase in the
calculated cross sections, the poles in the amplitude corresponding to
graph 1(b) are unphysical~\cite{wei88}.  These poles are just the pion
condensates seen in early calculations of the pion polarization potential 
in nuclei~\cite{ose83}. This effect arises because nucleon-hole and
$\Delta$-hole interactions are strongly attractive at short distances.
In actual situation, one should also include the strong short-ranged
repulsion which comes from the exchange of heavy mesons and other 
many body effects; when this is done, the spurious pole problem 
disappears. The distortion effects, which has not been considered
here, may also "soften" the poles.

In Fig.~4, we show the kaon angular distributions corresponding to 
various final hypernuclear states excited in this reaction. We have 
taken $g^\prime = 0.5$ throughout in this figure. In all the cases
the diagram 1(b) makes a dominant contribution to the cross sections.
Clearly, the cross sections are quite selective about the excited
hypernuclear state, being maximum for the state of largest orbital
angular momentum. This is due to the large momentum transfer involved
in this reaction. It may be noted that in each case the angular
distribution has a maximum around 10$^\circ$  and not at 0$^\circ$ 
as seen in previous non-relativistic calculations for this
reaction. This is the consequence of using Dirac spinors for 
the bound states. There are several maxima in the upper and lower
components of the momentum space bound spinors \cite{shy95} in the
region of large momentum transfers. Therefore, in the kaon angular
distribution the first maximum may shift to larger angles
reflecting the fact that the bound state wave functions show
diffractive structure at higher momentum transfers~\cite{shy95}.

In Fig.~5, we show the dependence of our calculated cross sections on 
pion self-energy.  It is interesting to note that this has a rather
large effect. We also note a surprisingly large effect on the 
short range correlation (expressed schematically by the Landau- Migdal 
parameter $g^\prime$). Similar results have also been reported in 
case of the $(p,\pi)$ reactions. However, more definite
statements about the usefulness of $(p,K^+)$ reactions in
probing the medium effects on the pion propagation in nuclei
must await the inclusions of distortions in the incident and outgoing
channels and exchange of heavy intermediate mesons ($\rho$ and $\omega$). 

The relative importance of the contribution of each intermediate
resonance to our reaction is shown in Fig.~5. It is
clear that contributions from the $N^*$(1710) resonance dominate the
total cross section at this beam energy. We also note that
the interference terms of the amplitudes corresponding to various
resonances are not negligible. It should be emphasized that we have 
no freedom in choosing the relative signs of the interference terms.
This result is similar to that observed in the study of the elementary
$pp \rightarrow p\Lambda K^+$ reaction~\cite{shy99} at this beam energy.

The cross sections near the peak are found to be about 0.1 nb/sr
for the excitation of the $^{41}\!\!\!_\Lambda$Ca($0d_{3/2}$)
hypernuclear state at the beam energy of 2.0 GeV. This is of the same 
order of magnitude as the upper limit of the experimental center of
mass cross section deduced for this reaction on a $^4$He target at the
same beam energy~\cite{kin98}. Nevertheless, the short-range correlations
driven by heavy meson exchange diagrams could enhance these
cross sections. Also use of the bound state wave functions calculated within
a microscopic theory like DDRH~\cite{kei00} is desirable. Although
the spinors for various states calculated with our well-depth 
search method are in good agreement with those of the DDRH theory, the
parameters of the potential well are not unique.
Inclusion of these effects in our TNM model is in progress. The distortion
effects in the incident and outgoing channels are also being taken into
account in much the same way as it has been done in the TNM calculations
of the $(p,\pi)$ reaction in Ref.~\cite{shy91}.

\section{Summary and Conclusions} 

In summary, we have made the first exploratory study of
the $A(p,K^+){_\Lambda}B$ reaction on a $^{40}$Ca target 
within a fully covariant general two-nucleon mechanism where 
in the initial collision of the incident proton with one
of the target nucleons (which is mediated by the one-pion exchange
mechanism), $N^*$(1710), $N^*$(1650), and $N^*$(1720) baryonic
resonances are excited which subsequently propagate
and decay into the relevant channels. Wave functions of baryonic 
bound states are obtained by solving the Dirac equation with appropriate
scalar and vector potentials. We have ignored, for the time being, the
distortion effects in the incident and the outgoing channels as our main
aim in this paper has been to establish a realistic two nucleon model for
this reaction which has not been done before.

In our model the $(p,K^+)$ reaction   
proceeds predominantly via excitation of the $N^*$(1710) resonant
state at the beam energy around 2 GeV.
We find that the nuclear medium corrections to the intermediate
pion propagator introduce large effects on the kaon differential 
cross sections. There is also the sensitivity of the cross sections
to the short-range correlation parameter $g^\prime$ in the pion self-energy.
Thus, $(p,K^+)$ reactions may provide an interesting 
tool to investigate the medium corrections on the pion propagation
in nuclei. A useful future study would be to investigate this
reaction also within an initial state kaon exchange model as it might 
make it possible to probe the kaon self energy in the
nuclear medium. Results of such calculations together with the distortion
effects will be presented elsewhere. 
 
This work has been supported by the Forschungszentrum J\"ulich.

\newpage
\begin{table}
\caption{Coupling constants and cutoff parameters for various vertices
used in the calculations.
}
\begin{ruledtabular}
\begin{tabular}{cccc}
Intermediate State & Decay channel & $g$ & Cutoff parameter\\ 
                   &               &     & \footnotesize {(GeV)} \\
\hline
$N$ & $N\pi$ & 12.56 & 1.2 \\
$N^*$(1710)& $N\pi$ & 1.04 & 1.2 \\
          & $\Lambda K^+$ & 6.12 & 1.2 \\
$N^*$(1650)& $N\pi$ & 0.81 & 1.2 \\
          & $\Lambda K^+$ & 0.76 & 1.2 \\
$N^*$(1720)& $N\pi$ & 0.21 & 1.2 \\
          & $\Lambda K^+$ & 0.87 & 1.2 \\
\hline
\end{tabular}
\end{ruledtabular}
\end{table}

\newpage
\begin{table}[here]
\caption  {Searched depths of vector and scalar potentials and the binding
energies of the $\Lambda$ and nucleon bound states. In each case we have
taken $r_v$, $r_s$ = 0.987 fm and $a_v$ = 0.68 fm and $a_s$ = 0.70 fm.}
\vspace{1.1cm}
\begin{tabular}{|c|c|c|c|} \hline
 State & Binding Energy & $V_v$ & $V_s$ \\
       &(\footnotesize{MeV})&(\footnotesize{MeV})&(\footnotesize{MeV})\\
\hline
$^{41}\!\!\!_\Lambda$Ca$(0s_{1/2})$ & 17.882 & 154.884 & -191.215  \\
$^{41}\!\!\!_\Lambda$Ca$(0p_{3/2})$ & 9.677  & 179.485 & -221.587  \\
$^{41}\!\!\!_\Lambda$Ca$(0p_{1/2})$ & 9.140  & 188.188 & -232.331  \\
$^{41}\!\!\!_\Lambda$Ca$(0d_{5/2})$ & 1.544  & 207.490 & -256.160  \\
$^{41}\!\!\!_\Lambda$Ca$(1s_{1/2})$ & 1.108  & 192.044 & -237.091  \\
$^{41}\!\!\!_\Lambda$Ca$(0d_{3/2})$ & 0.753  & 230.206 & -284.205  \\
$^{40}$Ca$(0d_{3/2})$ & 8.333  & 360.980 & -445.660  \\
\hline
\end{tabular}
\end{table}
\newpage
\begin{figure}
\includegraphics[width=0.6 \textwidth]{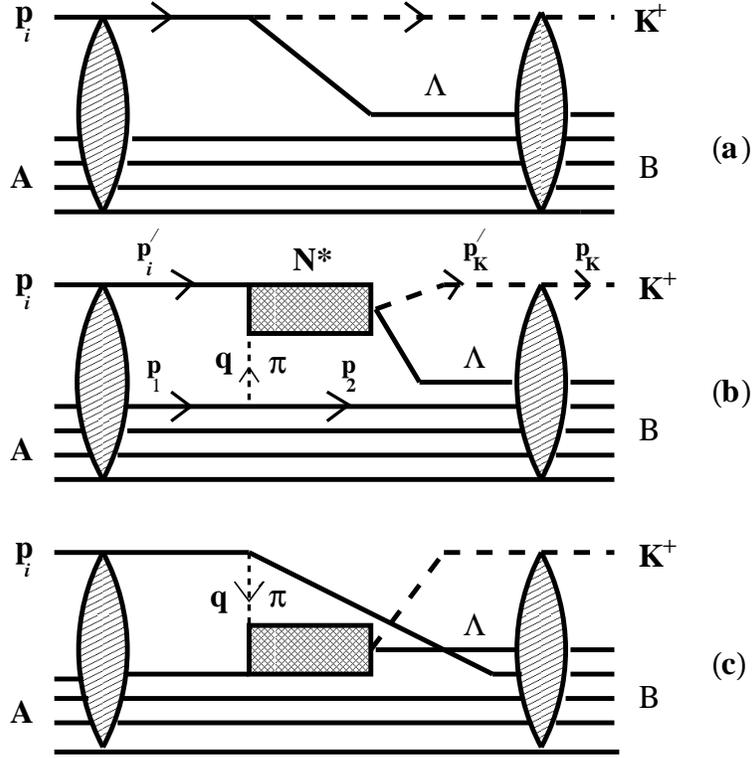}
\caption{Graphical representation of one- and two-nucleon models.
The elliptic shaded area represent the optical model interactions in 
the incoming and outgoing channels.  
}
\end{figure}
\newpage
\begin{figure}
\begin{center}
\includegraphics[width=0.6 \textwidth]{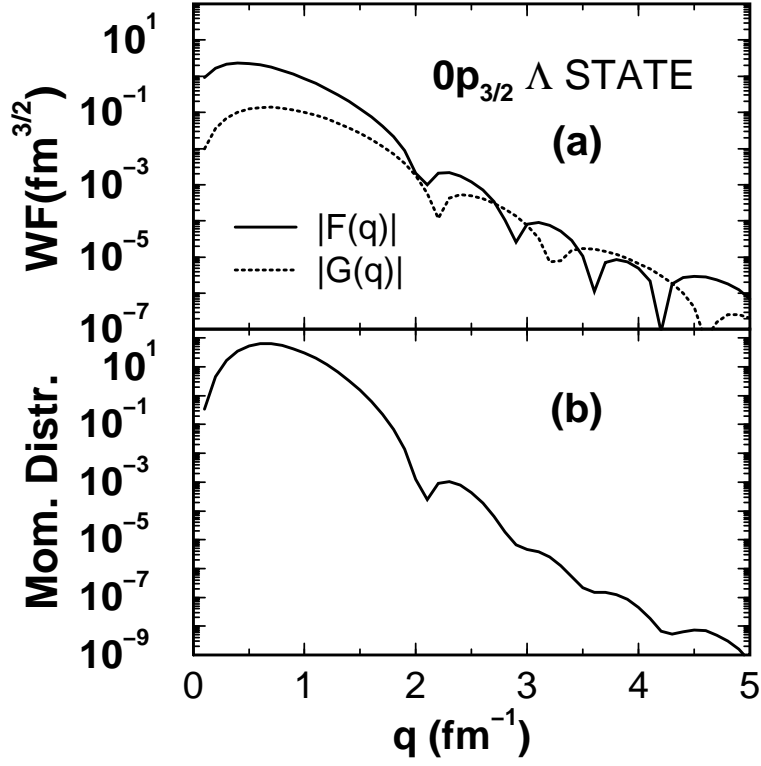}
\caption{ (a) Momentum space spinors (WF) for $0p_{3/2}$ $\Lambda$
orbit in $^{41}\!\!\!_\Lambda$Ca hypernucleus. $|F(q)|$ and $|G(q)|$ are
the upper and lower components of the spinor, respectively. (b)
momentum distribution (Mom. Distr.)for the same state hyperon calculted
with wave functions shown in (a).  
}
\end{center}
\end{figure}
\begin{figure}
\begin{center}
\includegraphics[width=0.6 \textwidth]{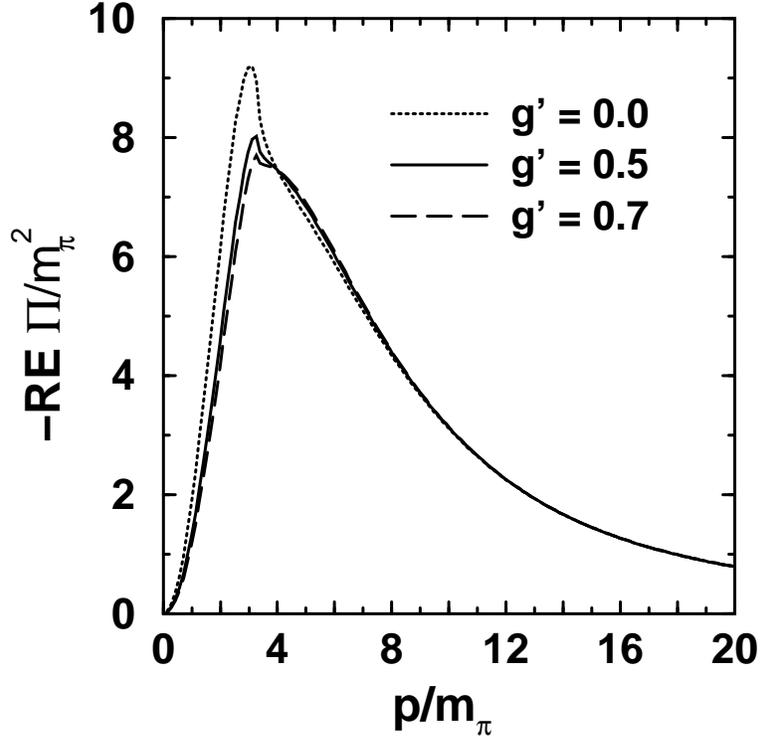}
\caption{
Real part of the renormalized pion self energy at zero energy as a 
function of momentum for symmetric nuclear matter at saturation 
density $\rho_0$. Contributions of both $ph$ and $\Delta h$ excitations
are included. The results obtained with $g'$ = 0.0, 0.5, and 0.7 are
shown by dotted, solid and dashed lines, respectively.
}
\end{center}
\end{figure}
\newpage
\begin{figure}
\begin{center}
\includegraphics[width=0.6 \textwidth]{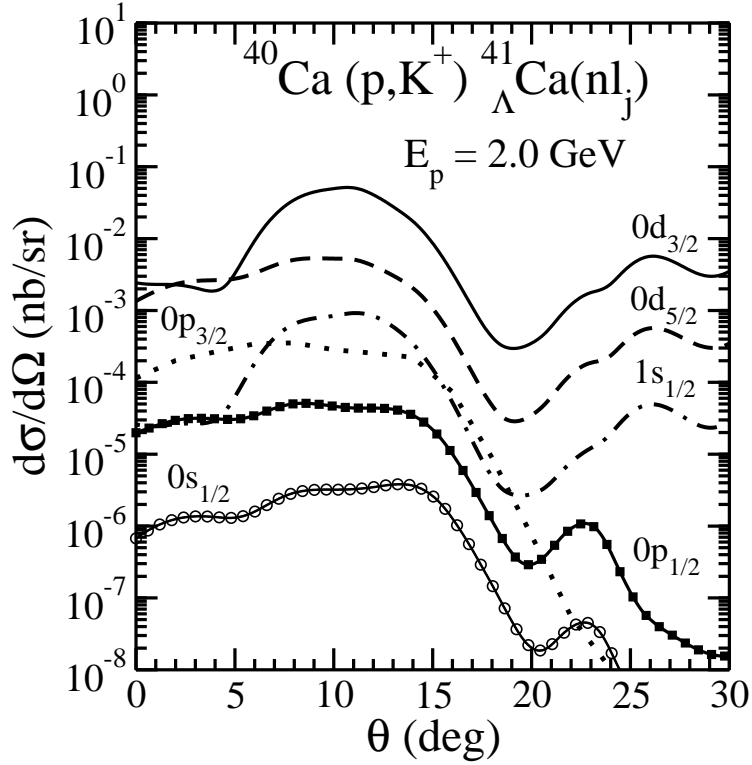}
\caption{Differential cross section for the 
$^{40}$Ca$(p,K^+)$$^{41}\!\!\!_\Lambda$Ca
reaction for the incident proton energy of 2.0 GeV for various
bound states of final hypernucleus as indicated in the figure.
The $\Lambda$ separation energies for $0d_{3/2}$, $0d_{5/2}$, $0p_{3/2}$,
$0p_{1/2}$, $1s_{1/2}$ and $0s_{1/2}$ states were taken to be 0.7529 MeV,
1.5435 MeV, 9.6768 MeV, 9.1400 MeV, 17.8820 MeV, and 1.1081 MeV,
respectively. The quantum number and the binding energy of the two
intermediate nucleon states were $0d_{3/2}$ and 8.3282 MeV, respectively.
}
\end{center}
\end{figure}
\newpage
\begin{figure}
\begin{center}
\includegraphics[width=0.6 \textwidth]{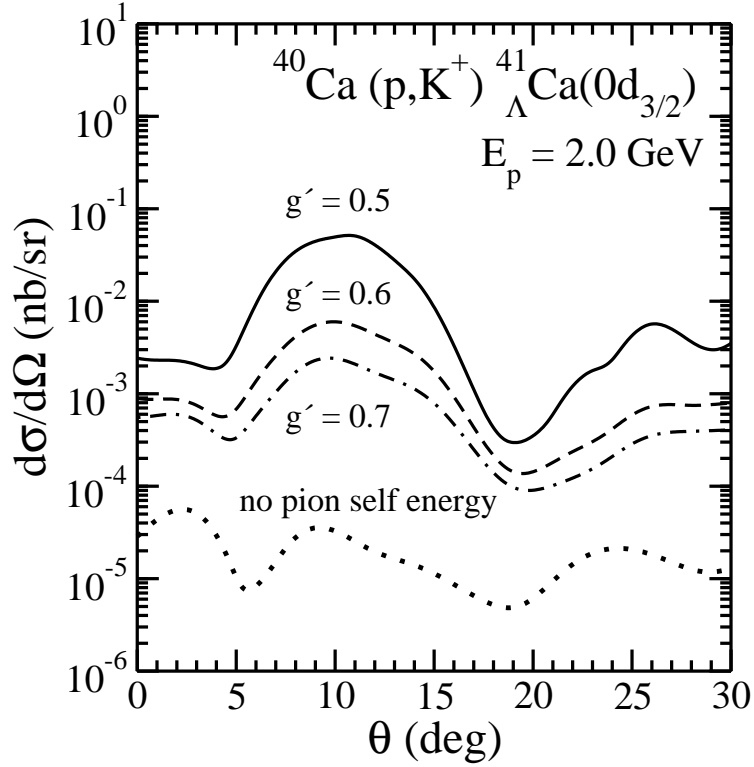}
\caption{Differential cross section for the 
$^{40}$Ca$(p,K^+)$$^{41}\!\!\!_\Lambda$Ca($0d_{3/2}$)
reaction for the incident proton energy of 2.0 GeV.  
The dotted line shows the results obtained without including the pion 
self-energy in the denominator of the pion propagator while
full, dashed and dashed-dotted lines represent the same calculated
with pion self-energy renormalized with Landau-Migdal parameter of
0.5,0.6 and 0.7, respectively.
}
\end{center}
\end{figure}
\newpage
\begin{figure}
\begin{center}
\includegraphics[width=0.6 \textwidth]{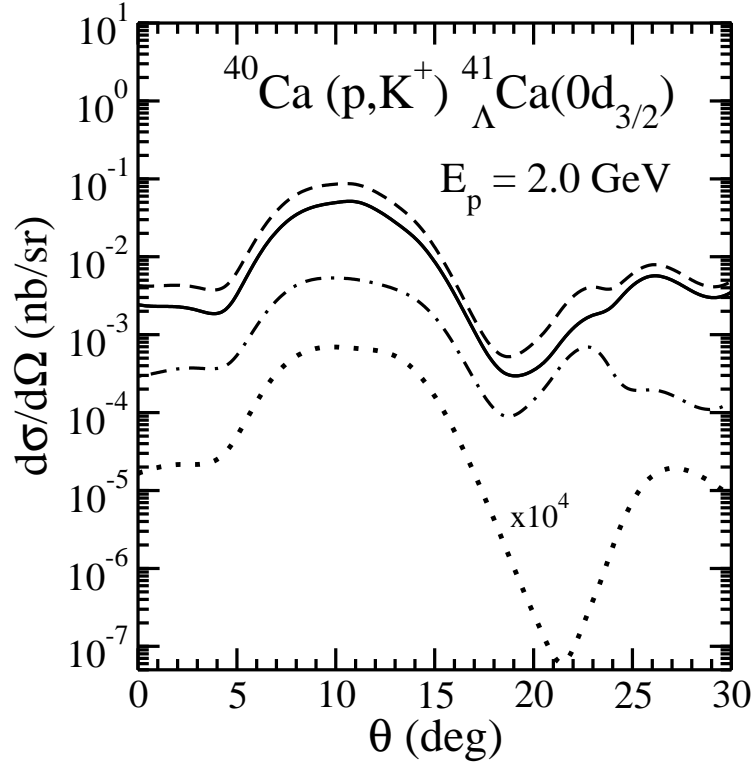}
\caption{Contributions of $N^*$(1710) (dashed line), $N^*$(1650)
(dashed-dotted line) and $N^*$(1720) (dotted line, plotted after
multiplying the actual cross sections by the factor shown) to the
differential cross section for the 
$^{40}$Ca$(p,K^+)$$^{41}\!\!\!_\Lambda$Ca$(0d_{3/2})$
reaction for the incident proton energy of 2.0 GeV.
Their coherent sum is shown by the full line. 
}
\end{center}
\end{figure}


\begin{thebibliography}{99}
\bibitem{chr89}
R. E. Chrien and C. B. Dover, Annu. Rev. Nucl. Part. Sci. {\bf39}, 113 (1989);
H. Kohri et al., Phys. Rev. C. {\bf 65}, 034607 (2002) and references therein.

\bibitem{pil91}
P. H. Pile et al., Phys. Rev. Lett. {\bf 66}, 2585 (1991).

\bibitem{hot01}
H. Hotchi et al., Phys. Rev. Lett. {\bf 64}, 044302 (2001).

\bibitem{hiy00}
E. Hiyama, M. Kamimura, T. Motoba, T. Yamada, and Y. Yamamoto, Phys. Rev. Lett.
{\bf 85}, 270 (2000);
H. Nemura, Y. Akaishi, and Y. Suzuki, Phys. Rev. Lett. {\bf 89}, 142504 (2002).
D. Vretenar, W. Poschl, G. A. Lalazissis, and P. Ring, Phys. Rev. C 
{\bf 57}, 1060 (1998).

\bibitem{kei00}
C. M. Keil, F. Hofmann, and H. Lenske, Phys. Rev. C {\bf 61}, 064309 (2000);
C. M. Keil and H. Lenske, Phys. Rev. C {\bf 66}, 054307 (2002).

\bibitem{kin98}
J. Kingler et al., Nucl. Phys. {\bf A634}, 325 (1998).

\bibitem{sch95}
O. B. W. Schult et al., Nucl. Phys. {\bf A585}, 247c (1995).

\bibitem{shi86}
S. Shinmura, Y. Akaishi, and H. Tanaka, Prog. Theo. Physik {\bf 76},
157 (1986).

\bibitem{kom95}
V. I. Komarov, A. V. Lago, and Yu. N. Uzikov, J. Phys. G: Nucl. Part. Phys.
{\bf 21}, L69 (1195);
V. N. Fetisov, Nucl. Phys. {\bf A639}, 177c (1998).

\bibitem{kri95}
B. V. Krippa, Z. Phys. {\bf A 351}, (1995) 411.

\bibitem{mea79}
D. F. Fearing and G. A. Miller, Annu. Rev. Nucl. Part. Sci. {\bf 29},
121 (1979).

\bibitem{coo82}
E. D. Cooper and H. S. Sherif, Phys. Rev. Lett. {\bf 47}, 818 (1982);
Phys. Rev. C {\bf 25}, 3024 (1982).

\bibitem{alo88}
P. W. F. Alons, R. D. Bent, J. S. Conte, and M. Dillig, Nucl. Phys. {\bf A480},
413 (1988).

\bibitem{pet98}
W. Peters, H. Lenske, and U. Mosel, Nucl. Phys. {\bf A640}, 89 (1998);
Nucl. Phys. {\bf A642}, 506 (1998).

\bibitem{shy95}
R. Shyam, W. Cassing and U. Mosel, Nucl. Phys. {\bf A586}, 557 (1995).

\bibitem{bar69}
M. V. Barnhill III, Nucl. Phys. {\bf A131}, 106 (1969);
L. D. Miller and H. J. Weber, Phys. Lett. {\bf 64B}, 279 (1976);
R. Brockmann and M. Dillig, Phys. Rev. C {15}, 361 (1977).

\bibitem{shy99}
R. Shyam, Phys. Rev. C {\bf 60}, 055213 (1999);
R. Shyam, G. Penner, and U. Mosel, Phys. Rev. C {\bf 63}, 022202(R) (2001).

\bibitem{kai99}
N. Kaiser, Eur. Phys. J. A {\bf 5}, 105 (1999).

\bibitem{gas00}
A. Gasparian, J. Heidenbauer, C, Hanhart, L. Kodratyuk, and J. Speth,
Phys. Lett. B {\bf 480}, 273 (2000).

\bibitem{lag91}
J. M. Laget, Phys. Lett. B {\bf 259}, 24 (1991).

\bibitem{bjo64}
J. D. Bjorken and S. D. Drell, {\it Relativistic Quantum Mechaniscs},
(McGraw-Hill, New York, 1964).

\bibitem{pen02}
G. Penner and U. Mosel, Phys. Rev. C {\bf 66}, 055211 (2002); {\bf 66},
055212 (2002).

\bibitem{lut03}
C. L. Korpa and M. F. M. Lutz, nucl-th/0306063.

\bibitem{mal02}
S. Mallik, Eur. Phys. J. C {\bf 24}, 143 (2002).

\bibitem{shy91}
R. Shyam, A. Engel, W. Cassing, and U. Mosel, Phys. Lett. {\bf B273},
26 (1991).

\bibitem{fru84}
S. Frullani and J. Mougey, Adv. Nucl. Phys. {\bf 14}, 1 (1984).

\bibitem{dmi85}
V. F. Dmitriev and Toru Suzuki, Nucl. Phys. {\bf A438}, 697 (1985).

\bibitem{jai88}
B.K. Jain,J. T. Londergan, and G. E. Walker, Phys. Rev. C {\bf 37}, 1564 
(1988).
 
\bibitem{her92}
T. Herbert, K. Wehrberger, and F. Beck, Nucl. Phys. {\bf A541}, 699 (1992).

\bibitem{wei88}
T. E. O. Ericson and W. Weise, {\it Pions and Nuclei}, (Clarendon,
Oxford, 1988).

\bibitem{ose83}
E. Oset, H. Toki, and W. Weise, Phys. Rep. {\bf 83},
282 (1982); G. E. Brown and W. Weise, Phys. Rep. 27, 1 (1976). 
\end{thebibliography}
\end{document}